\documentclass[12pt]{article}
\usepackage[centertags]{amsmath}
\usepackage{amssymb,amsfonts}
\usepackage{graphicx}
\usepackage{rotating}
\usepackage{latexsym}
\begin{document}

\title{Relativistic stars with polytropic equation of state}
\author{Sifiso A. Ngubelanga and Sunil D. Maharaj \\
 Astrophysics and Cosmology Research Unit
    School of Mathematics, \\Statistics and Computer Science
    University of KwaZulu-Natal\\
   Private Bag X54001
   Durban
   4000
   South Africa
}

\maketitle


\begin{abstract} 

 We consider the Einstein-Maxwell system of equations in the context of isotropic coordinates for matter distributions with anisotropy in the presence of an electric field. We assume a polytropic equation of state for the matter configuration. New classes of exact solutions are generated for different polytropic indices. The model is well behaved and we can regain masses of several observed objects, in particular we obtain the mass of the star PSR J1903+327. 
\end{abstract}

\section{Introduction}

Polytropic equations of state are important in a variety of astrophysical applications. Traditionally the polytropic equation of state has been used to describe a completely degenerate gas in Newtonian theory and special relativity, an isothermal ideal gas and a completely convective star. In particular situations the star may be highly compact so that it is necessary to incorporate the effects of general relativity. This would be true for a neutron star as an example. It is for this reason that polytropes in general relativity were first studied by Tooper \cite{Tooper, Tooper1, Tooper2}. Recently Herrera and Barreto \cite{Herrera} generated the general formalism for polytropic general relativistic stars with anisotropic pressure. There are several sources for anisotropy including intense magnetic fields, as in neutron stars, white dwarfs and magnetized strange quark matter, and viscosity. The particular case of a conformally flat spacetime for a polytrope with anisotropic matter was investigated by Herrera \textit{et} \textit{al} \cite{Herrera1}. They point out the intriguing possibility of applying these polytropic models to describe super-Chandrasekkar white dwarfs with masses above the Chandrasekkar limit. 

Exact solutions with a polytropic equation of state of the Einstein field equations are rare because of the high degree of nonlinearity present in the model. A recent class of exact solutions for charged anisotropic polytropic spheres was found by Mafa Takisa and Maharaj \cite{Mafa}. An interesting feature of this class is that it contains the stellar models of Feroze and Siddiqui \cite{Feroze} with a quadratic equation of state. Anisotropic polytropic spheres, in the absence of charge, have been analysed by Thirukkanesh and Ragel \cite{Thirukkanesh} who produced exact solutions for two polytropic indices. Stellar models with a modified Van der Waals equation of state, with a polytropic exponent for the matter distribution, were studied by Malaver \cite{Malaver, Malaver1} and Thirukkanesh and Ragel \cite{Thirukkanesh1}. They generated solutions in terms of elementary functions and related them to compact relativistic matter. 

Ngubelanga \textit{et} \textit{al} \cite{Ngubelanga, Ngubelanga1} found new solutions to the Einstein-Maxwell system with anisotropy and charge in isotropic coordinates. Their models satisfy a linear and quadratic equation of state, respectively. We show in this paper that its possible to extend their approach to a polytropic equation of state. Our aim is to describe charged anisotropic matter in isotropic coordinates with polytropic equation of state. In section 2, we present the Einstein-Maxwell system of equations, and we also introduce the polytropic equation of state. In section 3, we choose a form for one of the gravitational potentials and the electric field intensity in order to perform the integration. New classes of solutions are found for different polytropic indices. In section 4, some concluding remarks are made.

\section{Field equations}

We consider the metric for a static spherically fluid inside a general relativistic star of the form 

\begin{equation}
\label{eq:g1}  ds^{2} = -A^{2}(r)dt^{2} + B^{2}(r)[dr^{2} + r^{2} (d \theta^{2} + \sin^{2} \theta d \phi^2)],
\end{equation}

\noindent in isotropic coordinates. It is helpful to introduce new metric functions so that 

\begin{equation}
\label{eq:g2} x  \equiv  r^{2},	\hspace{0.5cm} L  \equiv  B^{-1},	  \hspace{0.5cm}   G \equiv  LA.
\end{equation}

\noindent Then (\ref{eq:g1}) becomes 

\begin{equation}
\label{eq:g3}		ds^2 = -\left(\frac{G}{L}\right)^2dt^2+L^{-2}\left[\left(\frac{1}{4x}\right)dx^2+x(d \theta^{2} + \sin^{2} \theta d \phi^2)\right],
\end{equation}

\noindent which helps to simplify the field equations. Since we are studying an anisotropic fluid distribution with charge we take the energy momentum tensor to be 

\begin{equation}
\label{eq:g4} 	T_{ij} = \mbox{diag}\left(-\rho-\frac{1}{2}E^2, p_{r}-\frac{1}{2}E^2,p_{t}+\frac{1}{2}E^2,p_{t}+\frac{1}{2}E^2\right),
\end{equation}

\noindent where $\rho$ is the energy density, $p_{r}$ is the radial pressure, $p_{t}$ is the tangential pressure and $E$ represents the electric field intensity. These matter variables are measured relative to the fluid four-velocity $u^{i}=\frac{L}{G} \delta^{i}_{0}$.

For the metric (\ref{eq:g1}) and the charged anisotropic matter distribution (\ref{eq:g4}), the Einstein-Maxwell equations can be written as 

\begin{subequations}	\label{eq:g5}
\begin{eqnarray}
\label{eq:g5a} 8 \pi  \rho + \frac{1}{2} E^{2}   &=& 4[2xLL_{xx}-3(xL_{x}-L)L_{x}],\\                                                        \nonumber\\
\label{eq:g5b} 8 \pi   p_{r}  -\frac{1}{2} E^{2}   &=& 4L(L-2xL_{x})\frac{G_{x}}{G}-4(2L-3xL_{x})L_{x},\\                                      \nonumber\\
\label{eq:g5c}  8 \pi  p_{t}   + \frac{1}{2} E^{2}  &=& 4xL^{2}\frac{G_{xx}}{G}+4L(L-2xL_{x})\frac{G_{x}}{G}-4(2L-3xL_{x})L_{x}-8xLL_{xx},\\                                                        \nonumber\\
\label{eq:g5d}  \sigma^2 &=& \frac{L^2  (E+xE_{x})^2}{4\pi^2 x} ,
\end{eqnarray}
\end{subequations}

\noindent where $\sigma$ is the proper charge density. Subscripts denote derivatives with respect to the coordinate $x$ and we use units where $c=1$ and $G=1$. We take the equation of state to be polytropic 

\begin{eqnarray}	
\label{eq:g6}			8 \pi p_{r} = \alpha \rho^{\Gamma},	           
\end{eqnarray}

\noindent where the adiabatic constant $\Gamma$ is related to the polytropic index $n$ by $\Gamma = 1+\frac{1}{n}$. Equations (\ref{eq:g5b}) and (\ref{eq:g5c}) give the equation of pressure anisotropy 

\begin{eqnarray}
\label{eq:g7} \frac{G_{xx}}{G}-2\frac{L_{xx}}{L} = \frac{(\Delta + E^2)}{4xL^2},
\end{eqnarray}

\noindent where $\Delta = 8\pi(p_{t} - p_{r})$ is the measure of anisotropy. The mass function can be written in the form 

\begin{equation}
\label{eq:g8}		m(x) = 2 \pi \int^{x}_{0}{\frac{1}{\sqrt{\omega}} \left[\omega \rho(\omega) + \frac{E^2}{8 \pi}\right] d\omega}.
\end{equation}

\noindent Then the Einstein-Maxwell system of equations can be written as

\begin{subequations}	\label{eq:g9}
\begin{eqnarray}
\label{eq:g9a}		8 \pi \rho		 &=& 4[2xLL_{xx}-3(xL_{x}-L)L_{x}]-\frac{1}{2}E^2,\\	\nonumber\\
\label{eq:g9b}		8 \pi p_{r}		 &=& \alpha \rho^{1+\frac{1}{n}},\\	\nonumber\\
\label{eq:g9c}		8 \pi p_{t}		 &=& 8 \pi p_{r}+\Delta,\\	\nonumber\\
\label{eq:g9d}		8 \pi \Delta		 &=&4xL^{2}\frac{G_{xx}}{G}+4L(L-2xL_{x})\frac{G_{x}}{G}-4(2L-3xL_{x})L_{x}-8xLL_{xx}-\frac{1}{2}E^2 	\nonumber\\	\nonumber\\
						& &-8\pi \alpha\left\{\frac{8[2xLL_{xx}-3(xL_{x}-L)L_{x}]-E^2}{16\pi}\right\}^{1+\frac{1}{n}} ,\\			\nonumber\\
\label{eq:g9e}		\frac{G_{x}}{G}	&=&\frac{2\pi \alpha}{L(L-2xL_{x})} \left\{\frac{8[2xLL_{xx}-3(xL_{x}-L)L_{x}]-E^2}{16\pi}\right\}^{1+\frac{1}{n}}\nonumber\\		\nonumber\\
						& & +\frac{(2L-3xL_{x})L_{x}}{L(L-2xL_{x})}-\frac{E^2}{8L(L-2xL_{x})},\\	\nonumber\\
\label{eq:g9f}		\sigma^2		&=& \frac{L^2(E+xE_{x})^2}{4\pi^2x},\\	\nonumber\\
\label{eq:g9g}			E^2 &=& x(c+dx).
\end{eqnarray}
\end{subequations}

\noindent The system of differential equations (\ref{eq:g9}) is nonlinear, however we can integrate it exactly for particular polytropic indices $n$ to produce models in terms of elementary functions.

\section{Polytropic models}

We need to integrate (\ref{eq:g9}) to obtain explicit forms for the gravitational potentials and the matter variables. To achieve this we select the potential $L$ and the electric field intensity $E$, respectively, as 

\begin{subequations}		\label{eq:g10}
\begin{eqnarray}
\label{eq:g10a}		L   &=& a+bx,\\ 
\label{eq:g10b}		E^2 &=& x(c+dx),
\end{eqnarray}
\end{subequations}

\noindent where $a$, $b$, $c$ and $d$ are constants. Similar choices were made by Ngubelanga $et$ $al$ \cite{Ngubelanga, Ngubelanga1} to generate models with linear and quadratic equations of state in isotropic coordinates. We expect the form in (\ref{eq:g10}) to also produce acceptable stellar configurations with the polytropic equation of state (\ref{eq:g6}) for particular indices $n$.

The fundamental equation to be solved is the nonlinear field equation (\ref{eq:g9e}). We can integrate this equation for the polytropic indices $n=1, \frac{1}{2}, 2, \frac{2}{3}$ to find the gravitational potential $G$. Consequently we can write the line element in the form 

\begin{eqnarray}
\label{eq:g11}		ds^2  &=&  -K(a-br^2)^{2\Psi} (a+br^2)^{2(\Phi -1)} e^{2N(r)}\nonumber\\	
						 & & +(a+br^2)^{-2}\left[dr^2+r^2(d \theta^2 +\sin^2 \theta d\phi^2)\right],
\end{eqnarray}

\noindent where the constants $\Psi$, $\Phi$, and the function $N(r)$ are given in Table \ref{tab:table1} for the different polytropic indices. Once (\ref{eq:g9e}) is integrated then the system of differential equations (\ref{eq:g9}) generates the quantities associated with the matter and charge.

\begin{table}
\caption{The metric constants $\Psi$, $\Phi$ and the function $N(r)$ for various polytropic indices.}
\begin{center}
\begin{tabular}{|l|l|}
\hline
 &  $n=1$ \\
\hline
$\Psi$ &  	$\frac{1}{256 \pi b^5} \{16 \pi b^2 [b(c-8b^2)+ad] - a \alpha [a^2d^2+b(c-24b^2)(2ad+bc-24b^3)]\}$    \\	
&\\
$\Phi$ &	$\frac{1}{256\pi b^5}\{16\pi b^2[b(24b^2+c)-ad]+a \alpha[a^2d^2-b(24b^2+c)(2ad-bc-24b^3)]\}$  \\	
&\\
$N(r)$ &	$\frac{r^2}{384 \pi b^4}\{48 \pi b^2d-3\alpha [b^2c^2+ad(ad-48b^3)]-b^2d\alpha (3c+dr^2)r^2\}$ \\			
&\\
 \hline
  &   $n=\frac{1}{2}$ \\
 \hline
 $\Psi$ &   $\frac{1}{4096 \pi^{2}b^7} \{256\pi^2 b^4[b(c-8b^2)+ad]+a^2\alpha[a^2d^2(ad+3bc-72b^3)$\\
		& 			$-3ab^2d(48b^2c-c^2-576b^4)-1728b^7(8b^2-c)-b^3c^2(72b^2-c)]\}$    \\
 &\\
$\Phi$ &	  $\frac{1}{4096\pi^2b^7}\{256\pi^2b^4(b(c+24b^2)-ad)-a^2\alpha[a^2d^2(ad-3bc-72b^3)$ \\
	   &	   $+3ab^2d(48b^2c+c^2+576b^4)-1728b^7(8b^2+c)-b^3c^2(72b^2+c)]\} $ \\
&\\
$N(r)$ &  $\frac{r^2}{122880 \pi^2 b^6}\{15360\pi^{2}b^4d+60a\alpha [a^3d^3-72a^2b^3d^2+1728ab^6d$\\
		&$+3ab^2c^2d-72b^5c^2]+30b^2c\alpha[3a^2d^2-144ab^3d+b^2c^2]r^2$\\
		& $+20b^2d\alpha [a^2d^2-72ab^3d+3b^2c^2]r^4+3b^4d^2\alpha(15c+4dr^2)r^6\}$  \\
 \hline
 &    $n=2$\\
 \hline
$\Psi$ &    $\frac{1}{512 \pi^\frac{1}{2} ab^4d^2} \{ 32 \pi^\frac{1}{2} abd^2[b(c-8b^2)+ad]+(-d)^\frac{1}{2}\alpha[4a^2d^2(2ad+3bc)+b^2c^2(6ad+bc)]\} $   \\
&\\
$\Phi$ &	$\frac{1}{512 \pi^\frac{1}{2} ab^4d^2} \{ 32 \pi^\frac{1}{2} abd^2[b(c+24b^2)-ad]+(-d)^\frac{1}{2}\alpha[4a^2d^2(2ad-3bc)+b^2c^2(6ad-bc)]\} $  \\
&\\
$N(r)$ &	$\frac{r^2}{64\pi^\frac{1}{2}b^2}[8\pi^\frac{1}{2}d+(-d)^\frac{1}{2}\alpha(3c+dr^2)]$  \\
 \hline
 &   $n=\frac{2}{3}$ \\
\hline
$\Psi$ & $ \frac{1}{32768\pi^\frac{3}{2}ab^6d^3}\{2048\pi^\frac{3}{2}ab^3d^3[ad+b(c-8b^2)]$\\
		& $-(-d)^\frac{1}{2}\alpha[16a^4d^4(5bc+2ad)+40a^2b^2c^2d^2(bc+2ad)+b^4c^4(bc+10ad)]\}$     \\
&\\
$\Phi$ &	$ \frac{1}{32768\pi^\frac{3}{2}ab^6d^3}\{2048\pi^\frac{3}{2}ab^3d^3[b(c+24b^2)-ad]$ \\
		& $+(-d)^\frac{1}{2}\alpha[16a^4d^4(5bc-2ad)+40a^2b^2c^2d^2(bc-2ad)+b^4c^4(bc-10ad)]\}$  \\
&\\
$N(r)$ &	$ \frac{r^2}{6144\pi^\frac{3}{2}b^4d}\{768\pi^\frac{3}{2}b^2d^2-(-d)^\frac{1}{2}\alpha[15c(2a^2d^2+b^2c^2)$\\				   		& $+3d(2a^2d^2+5b^2c^2)r^2+b^2d^2(10c+3dr^2)r^4]\}$   \\
\hline
\end{tabular}
\end{center}
\label{tab:table1}
\end{table}

Since the metric functions $G$ and $L$ in (\ref{eq:g11}) are now known explicitly, the matter and charge variables can be written completely in terms of elementary functions for the polytropic indices $n=1, \frac{1}{2}, 2, \frac{2}{3}$. The energy density is given by $8 \pi \rho = 12ab-\frac{1}{2}r^2(c+dr^2)$. The radial pressure then follows via the equation of state $p_{r} = \alpha \rho^{1+\frac{1}{n}}$. The degree of anisotropy (and hence the tangential pressure) is generated from (\ref{eq:g7}) which is a lengthy and complicated expression. The electric field intensity is the polynomial $E^2 = r^2(c+dr^2)$ and the proper charge density is given by $\sigma^2 = \frac{1}{16\pi^2(c+dr^2)}(a+br^2)^2(3c+4dr^2)^2$. The matter and charge variables are continuous and well behaved in the interior of the sphere. 

The exact solutions found in this paper may be used to study the physical features of a general relativistic stellar model. As an example we consider the polytropic index $n=1$ and the equation of state has the form $p_{r} = \alpha \rho^2$. Table \ref{tab:table2} indicates the behaviour of $\rho$, $p_{r}$, $p_{t}$, $\Delta$, $m$, $E^2$ and $\sigma$ in the interior of the body. The energy density and radial pressure are decreasing away from the centre. The tangential pressure and anisotropy are increasing functions and finite. Clearly the mass function increases with increasing radial values. The electric field intensity and charge density are regular in the interior. It is interesting to observe that we can obtain the mass $m=1.667 $ $M_\odot$ which corresponds to the star PSR J1903+327. We point out that this star also arises in the analysis of Ngubelanga $et$ $al$ \cite{Ngubelanga1} with a general quadratic equation of state. Other stellar masses can be generated for different parameter values than those used in Table \ref{tab:table2}. In addition polytropic indices may also be used to perform a physical analysis. A detailed analysis of the physical features and relationship to observed stellar objects will be carried out separately for the general polytrope. Our intention in this paper was to demonstrate the existence of polytropic solutions for the choice (\ref{eq:g10}).

\begin{table}
\caption{Variation of energy density, radial pressure, tangential pressure, measure of anisotropy, mass, electric field intensity and charge density for charged bodies from the centre to the surface with parameters $a=1.65143$, $b=0.504167$, $c=0.01$, $d=0.01$, $n=1$ and $\alpha =0.931$.}
\begin{center}
\begin{tabular}{|c|c|c|c|c|c|c|c|}
\hline
 $r$ & $\rho(r)$ & $p_{r}(r)$ & $p_{t}(r)$ & $\Delta(r)$  & $m$ & $E^2$ & $\sigma^2$\\
\hline
 0.1 & 0.397534 & 0.147129 & 0.147702 & 0.000573287 & 0.00166686 & 0.000101 & 0.00158992  \\
 0.2 & 0.397527 & 0.147124 & 0.159383 & 0.0122589   & 0.013335   & 0.000416 & 0.00169896  \\
 0.3 & 0.397516 & 0.147116 & 0.17938  & 0.0322645   & 0.0450063  & 0.000981 & 0.00188841  \\
 0.4 & 0.397499 & 0.147103 & 0.209308 & 0.0622052   & 0.106684   & 0.001856 & 0.00217005  \\
 0.5 & 0.397473 & 0.147084 & 0.25172  & 0.104636    & 0.20837    & 0.003125 & 0.00256092  \\
 0.6 & 0.397438 & 0.147058 & 0.310532 & 0.163474    & 0.360071   & 0.004896 & 0.00308389  \\
 0.7 & 0.39739  & 0.147023 & 0.391743 & 0.244721    & 0.571787   & 0.007301 & 0.00376848  \\
 0.8 & 0.397327 & 0.146976 & 0.504716 & 0.357741    & 0.853521   & 0.010496 & 0.00465182  \\
 0.9 & 0.397244 & 0.146914 & 0.664513 & 0.517599    & 1.21527    & 0.014661 & 0.00577994  \\
 1.  & 0.397138 & 0.146836 & 0.896413 & 0.749578    & 1.667      & 0.02     & 0.00720911  \\
\hline
\end{tabular}
\end{center}
\label{tab:table2}
\end{table}

\begin{table}
\caption{Masses of observed objects corresponding to the parameters $c=0.01$ and $d=0.01$.}
\begin{center}
\begin{tabular}{|c|c|c|c|c|c|c|c|}
\hline
 Star & Observed  & Parameter $a$  & Parameter $a$ & Parameter $a$ \\
	& mass $m$ $(M_\odot)$	& ($b=0.258$)	&($b=0.800$) &	($b=0.935$) \\
\hline
 PSR J1614-2230 	& 1.97 	& 3.81432 & 1.23012  & 1.05251	\\
 Vela X-1 			& 1.77 	& 3.42672 & 1.10512  & 0.945554 \\
 4U 1608-52 		& 1.74 	& 3.36858 & 1.08637  & 0.929511 \\
 PSR J1903+327 		& 1.667 & 3.22711 & 1.040745 & 0.890474 \\
 4U 1820-30 		& 1.58 	& 3.0585  & 0.98637  & 0.84395  \\
 Cen X-3 			& 1.49 	& 2.88409 & 0.930119 & 0.795822 \\
 EXO 1785-248 		& 1.3  	& 2.51588 & 0.81137  & 0.69422  \\
 SMC X-4 			& 1.29 	& 2.4965  & 0.80512  & 0.68887  \\
 LMC X-4 		   	& 1.04 	& 2.012   & 0.64887  & 0.55518  \\
 SAX J1808.4-3658 	& 0.9 	& 1.74068 & 0.561369 & 0.480316 \\
 4U 1538-52 		& 0.87 	& 1.68254 & 0.542619 & 0.464273 \\
 Her X-1  			& 0.85 	& 1.64378 & 0.530119 & 0.453578 \\
\hline
\end{tabular}
\end{center}
\label{tab:table3}
\end{table}

\section{Discussion}

It is important to note that solutions in this paper generate masses from (\ref{eq:g8}) which are physically reasonable. We demonstrate this by making suitable choices of the parameters $a$, $b$, $c$ and $d$. In Table \ref{tab:table3} we give twelve masses for recent observed compact objects. By varying the parameter $a$ in equation (\ref{eq:g8}), keeping the other parameters fixed at $c=0.01$, $d=0.01$ and setting the parameter $b$ to three different values ($b=0.258$, $b=0.800$ and $b=0.935$), we obtain the corresponding masses. These values are consistent with other investigations for the compact stars \cite{Abubekerov}-\cite{Rawls}. We point out that the observed masses range from $m=0.85$ $M_\odot$ to $m=1.97$ $M_\odot$ when we have set the value of $r=1$ (at the surface). Thus our approach allows for a wide range of masses for realistic astrophysical objects.

We have shown that the formalism of Ngubelanga \textit{et} \textit{al} \cite{Ngubelanga, Ngubelanga1} can be extended to a polytropic equation of state. Our aim in this paper was to obtain new polytropic exact solutions to the Einstein-Maxwell equations in isotropic coordinates for matter distributions with anisotropy in the presence of charge. We found new exact solutions for the polytropic indices $n=1, \frac{1}{2}, 2, \frac{2}{3}$. The classes of exact solutions to the Einstein-Maxwell systems are physical reasonable. In particular choosing the fixed parameters $a=1.65143$, $b=0.504167$, $c=0.01$, $d=0.01$ and $\alpha=0.931$ and using the solution for the polytropic index $n=1$, we regained the mass of the stellar object PSR J1903+327 which has a mass of $m=1.667 $ $M_\odot$. This indicates the approach followed in this paper is likely to produce other meaningful models with a polytropic equation of state. In spite of the nonlinearity arising from the relationship $p_{r} = \rho^{1+\frac{1}{n}}$ the field equations do admit exact solutions by making a choice similar to (\ref{eq:g10}).

\noindent {\bf Acknowledgements}\\

SAN thanks the National Research Foundation and the University of KwaZulu-Natal for financial support. SDM acknowledges that this work is based upon research supported by the South African Research Chair Initiative of the Department of Science and Technology and the National Research Foundation. DST-NRF Centre of Excellence in Mathematical and Statistical Sciences (CoE-MaSS). Opinions expressed and conclusions arrived at are those of the author and are not necessarily to be attributed to the CoE-MaSS.\\

\end{document}